\documentclass[aps,preprint, 11pt,nofootinbib]{revtex4}
\usepackage{graphics}
\usepackage{slashed}
\usepackage{amssymb}
\usepackage{lscape}
\usepackage{amsmath}
\usepackage{rotating}
\usepackage{graphicx}
\graphicspath{ {images/} }
\begin{document}
\title{Fermionic origin of dark energy in the inflationary universe from Unified Spinor Fields}
\author{$^{2}$ Marcos R. A. Arcod\'{\i}a\footnote{E-mail address: marcodia@mdp.edu.ar},  $^{1,2}$ Mauricio Bellini
\footnote{E-mail address: mbellini@mdp.edu.ar} }
\address{$^1$ Departamento de F\'isica, Facultad de Ciencias Exactas y
Naturales, Universidad Nacional de Mar del Plata, Funes 3350, C.P.
7600, Mar del Plata, Argentina.\\
$^2$ Instituto de Investigaciones F\'{\i}sicas de Mar del Plata (IFIMAR), \\
Consejo Nacional de Investigaciones Cient\'ificas y T\'ecnicas
(CONICET), Mar del Plata, Argentina.}

\begin{abstract}
In this work we explore the boundary conditions in the Einstein-Hilbert action, by considering a displacement from the Riemannian manifold to an extended one. The latter is characterized by including spinor fields into the quantum geometric description of a noncommutative spacetime. These fields are defined on the background spacetime, emerging from the expectation value of the quantum structure of spacetime generated by matrices that comply with a Clifford algebra. We demonstrate that spinor fields are candidate to describe all known interactions in physics, with gravitation included. In this framework we demonstrate that the cosmological constant $\Lambda$, is originated exclusively by massive fermion fields that would be the primordial components of dark energy, during the inflationary expansion of an universe that describes a de Sitter expansion.
\end{abstract}
\maketitle

\section{Introduction and motivation}

The theory of inflation\cite{infl,infl1,infl2} solves the curvature problem by producing a very tiny
spatial curvature at the onset of the radiation epoch taking place
right after inflation. The spatial curvature can well grow during
the decelerated phase of expansion but it will be always
subleading provided inflation lasted for sufficiently long time. A general prediction of cosmological inflation is the generation of quantum fluctuations of the inflaton field\cite{cm,kb,bcms}, and of primordial gravitational waves (GW)\cite{tt,ns,sil,sil1}. The gravitational waves detected so far come from astrophysical phenomena\cite{..,..1,..2,..3,..4,..5}, but have not yet been detected as a cosmic background of gravitational radiation. The detection of GW produced during inflation would be of great importance for the understanding and corroborating of an inflationary epoch during the early phase of the expansion of the universe\cite{mat}, but the problem is that it is very weak. In the standard single-field, slow-roll inflationary scenario the tensor fluctuations of the metric are characterized by a nearly scale-invariant power-spectrum on super-Hubble scales. During inflation the universe grows quasi-exponentially describing a de Sitter expansion which is driven by a scalar field in a classical description, but its origin is still unkown and should be the same that of the dark energy (or the cosmological parameter). All these theoretical descriptions of the physical nature that drives the expansion of the universe contituite the same physical problem and is a mistery that deserves study.

One of the central problems in contemporary theoretical physics is the unification of quantum field theory with general relativity in a theory that contain both. In previous works we have developed a pure geometric spinor field theory on an arbitrary curved background, which is considered a Riemannian manifold\cite{ab}. In that theory, the spinor field with components $\hat{\Psi}^{\alpha}$ is responsible for the displacement from a Riemann manifold to an extended manifold, and the covariant derivative of the metric tensor in the Riemannian background manifold is null: $\nabla_{\alpha} g_{\beta\theta}=0$. However, the extended covariant derivative on the extended manifold (we denote the covariant derivative on the extended manifold with a $ \|  $ ), is nonzero: $ g^{\alpha\beta}_{\quad \|\gamma} \neq 0$\cite{mb}. Furthermore, we consider the coupling of the spinor fields with the background and their self-interactions in a generic manner. The theory is worked in 8 dimensions, 4 of them related to the space-time coordinates ($x^{\mu}$), and the other 4 related to the inner space ($\phi^{\mu}$), described by compact coordinates. The former have the spin components as canonical momentums: ($s_{\mu}$).  To describe a non-commutative spacetime, we shall consider unit vectors are $4\times 4$-matrices: $\bar{\gamma}^{\alpha}$, that generate a globally hyperbolic spacetime. These matrices generate the background metric and we include the spinor information in the spacetime structure that can describe quantum effects in a relativistic framework: $\left\{\bar{\gamma}_{\alpha},\bar{\gamma}_{\beta}\right\}=2\,g_{\alpha\beta}\,\mathbb{I}_{4\times 4}$, where $\mathbb{I}_{4\times 4}$ is the identity $4\times 4$-matrix.

In this work we extend the study when boundary terms are considered in the variation of the Einstein-Hilbert action in general relativity. We explore the emergent quantum physical dynamics that is obtained when we make a displacement from a background Riemann manifold to an extended one, by extending the Unified Spinor Field (USF) formalism developed in previous works, in order to make an rigorous study of the nature of the flux through the 3D-gaussian hypersurface on which the non-metricity is nonzero. In Sect. (\ref{cos}) we introduce the theoretical expression for the cosmological constant by studying the flux through the gaussian hypersurface in the boundary conditions of the minimal action principle and we demonstrates that $\Lambda$ is originated exclusively by fermionic fields. In Sect. (\ref{st}) we describe the noncommutative quantum spacetime. In Sect. (\ref{qd}) we develop the quantum dynamics of the spinor field components $\hat{\Psi}_{\alpha}$ on the extended manifold.
In Sect. (\ref{fo}), we make the Fourier expansion of the spinor fields from the scalar flux that cross the 3D-gaussian hypersurface in the boundary conditions of the minimal action principle. In the Sect. (\ref{ex}) we calculate the value of the cosmological constant $\Lambda$ in a de Sitter expansion of the universe and we obtain the value of the mass parameter $m$. Finally, in Sect. (\ref{conc}) we develop some final comments.

\section{Boundary conditions, flux and origin of the cosmological constant}\label{cos}

On the other hand, the boundary conditions in the minimum action's principle is a very important issue which must be taken into account to develop a physical theory\cite{york,gh}. For simplicity we can consider this topic in a general Einstein-Hilbert (EH) action
${\cal I} = \int d^4 x\, \sqrt{-\hat{g}}\, \left[\frac{{R}}{2\kappa} + {{\cal L}}\right]$ . It is important to notice that, after variation, we obtain
\begin{equation}\label{delta}
\delta {\cal I} = \int d^4 x \sqrt{-g} \left[ \delta g^{\alpha\beta} \left( G_{\alpha\beta} + \kappa T_{\alpha\beta}\right)
+ g^{\alpha\beta} \delta R_{\alpha\beta} \right],
\end{equation}
where $\kappa=8\pi G/c^4$. Here, ${{T}}_{\alpha\beta}$ is the the background stress tensor
\begin{equation}\label{bt}
{{T}}_{\alpha\beta} =   2 \frac{\delta {{\cal L}}}{\delta g^{\mu\nu}}  - g_{\mu\nu} {{\cal L}},\\
\end{equation}
and ${{\cal L}}$ is the Lagrangian density that describes the background physical dynamics. The last term in (\ref{delta}) is very important because takes into account boundary conditions. When that quantity is zero, we obtain the well known Einstein's equations without cosmological constant. In the general case, the classical Einstein equations, with boundary conditions included, results to be
\begin{equation}
\bar{G}_{\alpha\beta} = - \kappa\, \bar{T}_{\alpha\beta}, \label{e1}
\end{equation}
with $\bar{G}_{\alpha\beta} = {G}_{\alpha\beta} - \Lambda(x)\, g_{\alpha\beta}$. It is well known that Heisenberg suggested an unified quantum field theory of a fundamental spinor field describing all matter fields in their interactions\cite{hei1,hei2}. In his theory the masses and interactions of particles are a consequence of a self-interaction term of the elementary spinor field. The fact that manifolds with no-Euclidean geometry can help uncover new features of quantum matter makes it desirable to create manifolds of controllable shape and to develop the capability to add in synthetic gauge fields\cite{hohuang}. In our case, the full connections will be defined by
\begin{equation}\label{conn}
\hat{\Gamma}^{\alpha}_{\beta\gamma} = {\tiny\left\{ \begin{array}{cc}  \alpha \, \\ \beta \, \gamma  \end{array} \right\}}+ \hat{\delta
\Gamma}^{\alpha}_{\beta\gamma}={\tiny\left\{ \begin{array}{cc}  \alpha \, \\ \beta \, \gamma  \end{array} \right\}}+ \epsilon\, \hat{\Psi}^{\alpha}\,g_{\beta\gamma},
\end{equation}
which is an extension of whole introduced in\cite{og,ab,mb}. Here, $\hat{\Psi}^{\alpha}$ are quantum spinor field components that represent rather bosons or fermions\footnote{If we consider the case of fermions, we obtain the quantization rules in agreement with a anti-commutative algebra that comes from the Pauli's exclusion principle
\begin{eqnarray*}
&& \left< B\left| \left\{\hat{\Psi}_{\mu}({\bf x}, {\bf \phi}), \hat{\Psi}_{\nu}({\bf x}', {\bf \phi}') \right\}\right|B \right>= \frac{1}{2}\,\,\frac{s^2}{\hbar^2}\, \left\{\hat{\gamma}_{\mu}, \hat{\gamma}_{\nu} \right\}\,\mathbb{I}_{4\times 4} \,  \sqrt{\frac{\eta}{g}}  \,\,\delta^{(4)} \left({\bf x} - {\bf x}'\right) \,\delta^{(4)} \left({\bf \phi} - {\bf \phi}'\right), \\
&&\left< B\left| \left[\hat{\Psi}_{\mu}({\bf x}, {\bf \phi}), \hat{\Psi}_{\nu}({\bf x}', {\bf \phi}') \right]\right|B \right> =0.
\end{eqnarray*}
where $\sqrt{\frac{\eta}{g}}$ is the squared root of the ratio between the determinant of the Minkowsky metric: $\eta_{\mu\nu}$ and the metric that describes the background: $g_{\mu\nu}$. This ratio describes the inverse of the relative volume of the background manifold. Furthermore, in the case of bosons, they describe the algebra
\begin{eqnarray*}
&& \left< B\left| \left[\hat{\Psi}_{\mu}({\bf x}, {\bf \phi}), \hat{\Psi}_{\nu}({\bf x}', {\bf \phi}') \right]\right|B \right>  = \frac{s^2}{2 \hbar^2 } \left[\hat\gamma_{\mu} , \hat\gamma_{\nu}\right] \,  \sqrt{\frac{\eta}{g}} \,\,\delta^{(4)} \left({\bf x} - {\bf x}'\right) \,\delta^{(4)} \left({\bf \phi} - {\bf \phi}'\right), \\
&& \left< B\left| \left\{\hat{\Psi}_{\mu}({\bf x}, {\bf \phi}), \hat{\Psi}_{\nu}({\bf x}', {\bf \phi}') \right\}\right|B \right> =0,
\end{eqnarray*}
for the spin value $s$ (integer for bosons and $(2n+1)/2$ for fermions).}, and $\epsilon$ is a parameter to be determined. In the classical General Relativity theory $\Lambda$ is an invariant on the Riemann manifold, but not on the extended manifold (\ref{conn}). In this framework, the flux $\Theta$, on the extended manifold is given by
\begin{equation}\label{bor}
\hat{\Theta} = g^{\alpha\beta} \hat{\delta R}_{\alpha\beta} =
\left[\hat{\delta W}^{\alpha}\right]_{||\alpha} - \left({ g}^{\alpha\epsilon} \right)_{||\epsilon}  \,\hat{\delta\Gamma}^{\beta}_{\alpha\beta} +
 ( {g}^{\alpha\beta} )_{||\epsilon}  \,\hat{\delta\Gamma}^{\epsilon}_{\alpha\beta},
\end{equation}
where $||$ denotes the covariant derivative on the extended manifold defined by (\ref{conn}), with self-interactions included and $\hat{\delta\Gamma}^{\epsilon}_{\alpha\beta}= \epsilon\, \hat{\Psi}^{\alpha}\,g_{\beta\gamma}$. These self-interactions will describe quantum properties of spacetime, such that we must require
\begin{equation}\label{trans}
\bar{G}_{\alpha\beta} = {G}_{\alpha\beta} - {\Lambda}(x)\, g_{\alpha\beta}=-\kappa\, \bar{T}_{\alpha\beta},
\end{equation}
where
\begin{eqnarray}
\Theta&=&g^{\alpha\beta} \delta R_{\alpha\beta} = g^{\alpha\beta} \,\int \,dv'\,\left<B\left|\hat{\delta R}_{\alpha\beta}\right|B\right>, \label{r0} \\
\Lambda &=& \int \,dv'\,\left<B\left|\hat{\Lambda}\right|B\right>, \label{r1} \\
\bar{G}_{\alpha\beta} & = &{{G}}_{\alpha\beta}+ \int \,dv'\,\left<B\left|\hat{{\delta G}}_{\alpha\beta}\right|B\right>, \label{r2} \\
\bar{T}_{\alpha\beta} & = & {{T}}_{\alpha\beta}+ \int \,dv'\,\left<B\left| \hat{{\delta T}}_{\alpha\beta}\right|B\right>, \label{r3}
\end{eqnarray}
where $v'$ is the volume of the spacetime coordinated and of the inner space [which we shall study with more detail later, in Sect. \ref{cos}], on which we shall describe the spinor fields operators. In order to the new Einstein's equations to be fulfilled, we must require that:
\begin{eqnarray}\label{tran}
{{G}}_{\alpha\beta}& = &- \kappa\,{{T}}_{\alpha\beta},\label{tran1} \\
\hat{{\delta G}}_{\alpha\beta} &\equiv &- \hat{\Lambda}\, g_{\alpha\beta}=-\kappa\, \hat{{\delta T}}_{\alpha\beta}. \label{tran2}
\end{eqnarray}
Here, the equation (\ref{tran1}) describes the background (classical equations) without cosmlogoical parameter, and (\ref{tran2}) describes the new physics due to the flux that cross the 3D-gaussian hypersurface, which we shall consider that in this work is of quantum nature. Furthermore, the quantum stress tensor $ \hat{{\delta T}}_{\alpha\beta}$, is defined by
\begin{equation}
\hat{{\delta T}}_{\alpha\beta}= 2 \frac{\delta \hat{{\cal L}}}{\delta g^{\mu\nu}}  - g_{\mu\nu} \hat{{\cal L}}, \label{t2}
\end{equation}
where $\hat{{\cal L}}$ is the Lagrangian density which describes the quantum dynamics related to the flux that cross the 3D-gaussian hypersurface and must be the responsible for the existence of the quantum operator $\hat{\Lambda}$ that originates the cosmological parameter ${\Lambda}(x)$.

We consider the extended flux through the 3D-Gaussian hypersurface described by (\ref{bor}). From the equations (\ref{trans}) and
(\ref{Gab}), we can write
\begin{equation}\label{tra}
\bar{G}_{\alpha\beta} = {G}_{\alpha\beta} - {\Lambda}\, g_{\alpha\beta}= {G}_{\alpha\beta}+\delta G_{\alpha\beta},
\end{equation}
such that, we obtain
\begin{equation}
g^{\alpha\beta} \delta G_{\alpha\beta} =  \delta G = -4\Lambda,
\end{equation}
where
\begin{equation}
\delta G= g^{\alpha\beta} \delta G_{\alpha\beta} = \,\int\, d^4x\,\int\,d^4\phi\,\sqrt{-g}\,\left<B| g^{\alpha\beta}\hat{\delta G}_{\alpha\beta}|B\right>.
\end{equation}
Hence, we promote as a quantum operator $\hat{\Lambda}$, with expectation value $\Lambda = -{1\over 4} \delta G={1\over 4} \delta T$:
\begin{eqnarray}
\Lambda &=&\,\int\, d^4x\,\int\,d^4\phi\,\sqrt{-g}\,\left< B\left|\hat{\Lambda}  \right| B \right> , \nonumber \\
&=& \frac{1}{4} \,\int\, d^4x\,\int\,d^4\phi\,\sqrt{-g}\, \left< B\left|\hat{ \delta T} \right| B \right> ,\label{lam}
\end{eqnarray}
where $\hat{ \delta T}=  g^{\alpha\beta}\hat{\delta T}_{\alpha\beta}$. In Sect. \ref{ex} we shall calculate $\Lambda$ in a cosmological de Sitter expansion that describes the inflationary expansion of the primordial universe, in order to estimate its value due to primordial fermionic fields.

\section{Quantum structure of a noncommutative spacetime}\label{st}

In order to describe the dynamics of the operators $\hat{\Psi}_{\alpha}$ we must take into account the quantum structure of spacetime. To propose a description we shall consider that this spacetime is generated by a base of $4\times 4$-matrices, $\hat{\gamma}^{\alpha}$. The $\hat{\gamma}_{\alpha}= E^{\mu}_{\alpha} \gamma_{\mu}$ matrices which generate the background metric are related by the vielbein $E^{\mu}_{\alpha}$ to basis $\gamma_{\mu}$ in the Minkowski spacetime (in cartesian coordinates). By introducing these matrices we aim to include the spinor information in the spacetime structure and construct a non-commutative basis that can describe quantum effects in a relativistic framework. The Dirac and Majorana matrices are good candidates, but in general it is possible to use any basis that describe a globally hyperbolic spacetime, which is the global geometry necessary to obtain relativistic causality. It is expected that the background spacetime can emerge from the expectation value of a quantum structure of spacetime. We consider the variation of the quantum operator $\hat{X}^{\mu}$, which can be represented as
 \begin{equation}
\delta \hat{ X}^{\alpha}(x^{\nu}) = \frac{1}{(2\pi)^{2}} \int d^4 k \, \hat{\gamma}^{\alpha} \left[ b_k \, \hat{X}_k(x^{\nu}) + b^{\dagger}_k \, \hat{X}^*_k(x^{\nu})\right],
\end{equation}
such that $b^{\dagger}_k$ and $b_k$ are the creation and annihilation operators of spacetime, with $\left< B \left| \left[b_k,b^{\dagger}_{k'}\right]\right| B  \right> = \delta^{(4)}(\vec{k}-\vec{k'})$. We shall use the Heisenberg representation of the states $\left|B\right>$, in which the states do not evolve over time, but yes do it the operators. In an analogous manner we introduce the variation of the quantum operator $\hat{\Phi}^{\mu}$:
related to spin
\begin{equation}
\delta \hat{\Phi}^{\alpha}(\phi^{\nu}) = \frac{1}{(2\pi)^{2}} \int d^4 s \, \hat{\gamma}^{\alpha} \,\left[ c_s \, \hat{\Phi}_s(\phi^{\nu}) + c^{\dagger}_s \, \hat{\Phi}^*_s(\phi^{\nu})\right],
\end{equation}
where  $\left< B \left| \left[c_s,c^{\dagger}_{s'}\right]\right| B  \right> = \delta^{(4)}(\vec{s}-\vec{s'})$. In our case the background quantum state can be represented in a ordinary Fock space in contrast with Loop Quantum Gravity (LQG)\cite{aa,aa1}, where operators are qualitatively different
from the standard quantization of gauge fields. The variations and differentials of the operators $\hat{X}^{\mu}$ and $\hat{\Phi}^{\mu}$ on the extended Weylian manifold, are given respectively by
\begin{eqnarray}
\delta\hat{X}^{\mu}\left| B\right> &=& \left(\hat{X}^{\mu}\right)_{\|\alpha} dx^{\alpha}\left| B\right>, \qquad \delta\hat{\Phi}^{\mu} \left| B\right>= \left(\hat{\Phi}^{\mu}\right)_{\|\alpha} d\phi^{\alpha}\left| B\right>, \\
d\hat{X}^{\mu} \left| B\right>&=& \left(\hat{X}^{\mu}\right)_{,\alpha} dx^{\alpha}\left| B\right>, \qquad d\hat{\Phi}^{\mu} \left| B\right>= \left(\hat{\Phi}^{\mu}\right)_{,\alpha} d\phi^{\alpha}\left| B\right>,
\end{eqnarray}
where the covariant derivatives take into account the interaction of $\hat{X}^{\mu}$, with the geometrical spinor components $\hat{\Psi}^{\alpha}$:
\begin{eqnarray}
\left(\hat{X}^{\mu}\right)_{\|\beta}\left| B\right> &=& \left[\nabla_{\beta} \hat{X}^{\mu} + \epsilon\left(\hat{\Psi}^{\mu}  \hat{X}_{\beta} - \hat{X}^{\mu} \hat{\Psi}_{\beta}\right)\right]\left| B\right>, \\
\left(\hat{\Phi}^{\mu}\right)_{\|\beta}\left| B\right> &=& \left[\nabla_{\beta} \hat{\Phi}^{\mu} + \epsilon\left(\hat{\Psi}^{\mu}  \hat{\Phi}_{\beta} - \hat{\Phi}^{\mu} \hat{\Psi}_{\beta}\right)\right]\left| B\right>.
\end{eqnarray}
In order to recover a background theory in agreement with General Relativity, we must require that
the operators can be applied to some background  quantum state in the background curved space time, and they comply with
\begin{equation}\label{dif}
\delta\hat{X}^{\mu}\left|B\right> = dx^{\mu}\left|B\right>, \qquad \delta\hat{\Phi}^{\mu}\left|B\right> = d\phi^{\mu}\left|B\right>,
\end{equation}
where $\phi^{\alpha}$ are the four compact dimensions related to their canonical momentum components $s^{\alpha}$ that describe the spin.
Hence, the requisites for the equations (\ref{dif}) to be fulfilled, are
\begin{eqnarray}
 \epsilon\left( \hat{X}^{\mu} \hat\Psi_{\beta} -\hat{\Psi}^{\mu} \hat{X}_{\beta}\right)\left|B\right>  & = & \left\{ \begin{array}{cc}  \mu \, \\ \nu \, \beta  \end{array} \right\} \hat{X}^{\nu} \left|B\right>, \\
 \epsilon \left( \hat{X}_{\mu} \hat\Psi^{\alpha} -\delta^{\alpha}_{\mu} \hat{\Psi}^{\nu} \hat{X}_{\nu}\right)\left|B\right>
& = & g^{\alpha\beta} \left\{ \begin{array}{cc}  \nu \, \\ \beta \, \mu  \end{array} \right\} \hat{X}_{\nu} \left|B\right>, \\
 \epsilon \left(  \hat{\Phi}^{\mu} \hat\Psi_{\beta} -\hat{\Psi}^{\mu} \hat{\Phi}_{\beta} \right)\left|B\right>
& = & \left\{ \begin{array}{cc}  \mu \, \\ \nu \, \beta  \end{array} \right\} \hat{\Phi}^{\nu} \left|B\right>, \\
\epsilon \left(\hat{\Phi}_{\mu} \hat\Psi^{\alpha} -\delta^{\alpha}_{\mu} \hat{\Psi}^{\nu} \hat{\Phi}_{\nu}\right)\left|B\right> & = & g^{\alpha\beta} \left\{ \begin{array}{cc}  \nu \, \\ \beta \, \mu  \end{array} \right\} \hat{\Phi}_{\nu} \left|B\right>.
\end{eqnarray}

The squared norm of the bi-vectorial space of $\hat{\delta\Phi}$, and the inner product of $\hat{\delta X}$, are
\begin{equation}
\underleftrightarrow{\delta\Phi} \overleftrightarrow{\delta\Phi} \equiv\frac{1}{4} \left( \hat{\delta\Phi}_{\mu} \hat{\delta\Phi}_{\nu} \right) \left( \hat{\gamma}^{\mu} \hat{\gamma}^{\nu}\right), \qquad
\underrightarrow{\delta{X}} \overrightarrow{\delta{X}} \equiv  \frac{1}{4} \hat{\delta{X}}_{\alpha} \hat{\delta{X}}^{\alpha}.
\end{equation}
where $\hat{\Phi}^{\alpha}=\phi \,\hat{\gamma}^{\alpha} $, $\hat{X}^{\alpha}=x\, \hat{\gamma}^{\alpha} $ are respectively the components of the inner space and the coordinate space, and $\left(\hat{\gamma}_{\alpha}\hat{\gamma}_{\beta}\right)=\frac{1}{2} \left\{\hat{\gamma}_{\alpha}, \hat{\gamma}_{\beta}\right\}+\frac{1}{2} \left[\hat{\gamma}_{\alpha}, \hat{\gamma}_{\beta}\right]$, so that $\left(\hat{\gamma}^{\alpha}\hat{\gamma}^{\beta}\right)=\frac{1}{2} \left\{\hat{\gamma}^{\alpha}, \hat{\gamma}^{\beta}\right\}-\frac{1}{2} \left[\hat{\gamma}^{\alpha}, \hat{\gamma}^{\beta}\right]$, in order to obtain the relevant invariant: $\left(\hat{\gamma}_{\alpha}\hat{\gamma}_{\beta}\right)\left(\hat{\gamma}^{\alpha}\hat{\gamma}^{\beta}\right)=4\,\mathbb{I}_{4\times 4}$. The matrices $\hat\gamma^{\mu}$, comply with the Clifford algebra:
\begin{equation}
\hat{\gamma}^{\mu} = \frac{\bf{I}}{3!}\,\epsilon^{\mu}_{\,\,\alpha\beta\nu} \hat{\gamma}^{\alpha}\hat{\gamma}^{\beta}  \hat{\gamma}^{\nu} , \qquad \left\{\hat{\gamma}^{\mu}, \hat{\gamma}^{\nu}\right\} =
2 g^{\mu\nu} \,\mathbb{I}_{4\times 4}, \nonumber
\end{equation}
where ${\bf{I}}={\gamma}^{0}{\gamma}^{1}{\gamma}^{2}{\gamma}^{3}$ is the pseudoscalar, $\mathbb{I}_{4\times 4}$ is the identity matrix, and we define $\noindent{{\epsilon}^{\mu}_{\alpha\beta\nu}=g^{\mu\rho}\epsilon_{\rho\alpha\beta\nu}}$, with:
\begin{equation}
\epsilon_{\rho\alpha\beta\nu}=
\begin{cases}
1 &  \text{if} \ \rho\alpha\beta\nu \ \text{is an even permutation of}\ 0123 \\
-1 & \text{if} \ \rho\alpha\beta\nu \ \text{is an odd permutation of}\ 0123 \\
0 & \text{in any other case},
\end{cases}  \ \ .  \nonumber
\end{equation}
We shall consider the Weyl representation of matrices in cartesian coordinates:
\begin{eqnarray}
&& \gamma^0= \,\left(\begin{array}{ll}  0 & \mathbb{I} \\
\mathbb{I}  &  0 \ \end{array} \right),\qquad
\gamma^1=  \left(\begin{array}{ll} 0 &  -\sigma^1 \\
\sigma^1 & 0  \end{array} \right),  \nonumber \\
&& \gamma^2= \left(\begin{array}{ll} 0 &  -\sigma^2 \\
\sigma^2 & 0  \end{array} \right),  \qquad \gamma^3= \left(\begin{array}{ll} 0 &  -\sigma^3 \\
\sigma^3 & 0  \end{array} \right),\label{gamm}
\end{eqnarray}
where the Pauli matrices are
\begin{eqnarray}
&& \sigma^1 = \left(\begin{array}{ll} 0 & 1 \\
1  & 0  \end{array} \right), \quad \sigma^2 = \left(\begin{array}{ll} 0 & -i \\
i  & 0  \end{array} \right), \quad \sigma^3 = \left(\begin{array}{ll} 1 & 0 \\
0  & -1  \end{array} \right). \nonumber
\end{eqnarray}
The idea of introducing these matrices is to generate a globally hyperbolic (and non-commutative) spacetime where those matrices are a generalization of the unitary vectors. Once defined the spacetime on the desired global topology, one can describe the $4$-quantum spinor fields on this spacetime, in order for describe quantum effects in a relativistic framework. In other words, spacetime and matter are described by the same tetra-vectors $\hat{\gamma}^{\mu}$. The Dirac and Majorana matrices are good candidates, but in general it is possible to use any basis that describe a globally hyperbolic spacetime, which is the global geometry necessary to obtain relativistic causality.

The line elements to describe both, the coordinate spacetime and the inner spacetime, are
\begin{equation}\label{line}
dx^2 \delta_{BB'}  = \left<B\right| \underrightarrow{\delta{X}} \overrightarrow{\delta{X}} \left| B'\right> , \qquad  d\phi^2 \delta_{BB'}=
\left<B\right| \underleftrightarrow{\delta\Phi} \overleftrightarrow{\delta\Phi}\left| B'\right>.
\end{equation}
Notice that the coordinated spacetime takes into account only symmetrized contributions with respect to the matrices product, but the inner line element becomes from a bi-vectorial product that takes into account symmetric and anti-symmetric contributions of matrices.
The bi-vectorial squared norm of the spinor $\hat{S}$, is
\begin{equation}
\left\|\hat{S} \right\|^2 =  \left<B\left|\underleftrightarrow{S} \overleftrightarrow{S}\right|B\right> = \frac{1}{4} \left<B\left| \left( \hat{S}_{\mu} \hat{S}_{\nu} \right) \left( \hat{\gamma}^{\mu} \hat{\gamma}^{\nu}\right) \right|B\right> =s^2
\mathbb{I}_{4\times 4},
\end{equation}
where $\hat{S}_{\mu} = s\hat{\gamma}_{\mu}$. Each component of spin $\hat{S}_{\mu}$, is defined as the canonical momentum corresponding to the inner coordinate $\hat{\Phi}^{\mu}$, such that one can define an universal bi-vectorial invariant:
\begin{equation}\label{invariant}
\left<B\left| \underleftrightarrow{S} \overleftrightarrow{\Phi}\right|B\right> = \frac{1}{4} \left<B\left|\left( \hat{S}_{\mu} \hat{\Phi}_{\nu} \right) \left( \hat{\gamma}^{\mu} \hat{\gamma}^{\nu}\right)\right|B\right> =s \phi \,
\mathbb{I}_{4\times 4} = (2\pi n \hbar) \,\mathbb{I}_{4\times 4},
\end{equation}
with $n$-integer. In this framework, gravitons (which have $s=2\hbar$), will be invariant under $\phi=n\,\pi$ rotations, vectorial bosons (with $s =\hbar$), will be invariant under $\phi= 2n\pi$) rotations, fermions with $s =\frac{1}{2} \hbar$ will be invariant under $\phi= 4 n \pi$ rotations, meanwhile fermions with $s =\frac{3}{2} \hbar$ are invariant under $\phi= \frac{4}{3} n \pi$ rotations.

We define the variation of the metric tensor on the extended manifold, with respect to the background curved (Riemannian) one
\begin{equation}
\hat{g}_{\beta\alpha\|\gamma} \,\hat{\delta X}^{\gamma} \left| B\right> = \delta g_{\beta\alpha} \left| B\right>.
\end{equation}

\section{Fourier expansion of the flux $\hat{\Theta}$ and gauge-invariance}\label{fo}

The flux through the 3D-gaussian hypersurface, will be
\begin{equation}
\hat\Theta\left(x^{\mu}|\phi^{\nu}\right)= g^{\alpha\beta}\,\hat{\delta R}_{\alpha\beta} = -3 \epsilon \nabla_{\mu}\,\hat{\Psi}^{\mu},
\end{equation}
which means that the flux $\hat\Theta\left(x^{\mu}|\phi^{\nu}\right)$ is alway integrable on the background (Riemannian) manifold, and $\hat{\Psi}^{\mu}$ is invariant under transformations: $\hat{\tilde{\Psi}}_{\mu}=\hat{\Psi}_{\mu}-\nabla_{\mu} \hat{\Theta}$, with
\begin{equation}\label{dal}
\Box \hat{\Theta} =0,
\end{equation}
in order to $\nabla_{\mu}\hat{\tilde{\Psi}}^{\mu} = \nabla_{\mu}\hat{{\Psi}}^{\mu}$.

The flux $\hat{\Theta}$, of $\hat{\Psi}^{\alpha}$-field through the 3D-Gaussian hypersurface, can be represented according to (\ref{line}), as a Fourier expansion in the momentum-space (the asterisk denotes the complex conjugate):
\begin{eqnarray}
\hat{\Theta}\left(x^{\beta}|\phi^{\nu}\right)= \frac{1}{(2\pi)^4} \int d^4k \int d^4 s
\, \left[ A_{s,k}\, \Theta_{k,s}(x^{\beta}) e^{\frac{i}{\hbar} \underleftrightarrow{S} \overleftrightarrow{\Phi}}
+ B^{\dagger}_{k,s} \, \Theta^*_{k,s}(x^{\beta}) e^{-\frac{i}{\hbar} \underleftrightarrow{S} \overleftrightarrow{\Phi}}\right]. \nonumber
\end{eqnarray}

\subsection{Definition of $ \hat{\Psi}_{\alpha}$}

Therefore we can define the spinor $4$-vector components $\hat{\Psi}_{\alpha}\left(x^{\beta}|\phi^{\nu}\right)=\frac{\hat{\delta\Theta}}{\hat{\delta{\Phi}}^{\alpha}}$, in the momentum-space associated to (\ref{line}):
\begin{eqnarray}
 \hat{\Psi}_{\alpha}\left(x^{\beta}|\phi^{\nu}\right)&=& \frac{i}{\hbar (2\pi)^4} \int d^4k \int d^4s \frac{\delta \left(\underleftrightarrow{S} \overleftrightarrow{\Phi}\right)}{\hat{\delta\Phi}^{\alpha}} \left[ A_{s,k}\, \Theta_{k,s} e^{\frac{i}{\hbar} \underleftrightarrow{S} \overleftrightarrow{\Phi}} \right.-\left. B^{\dagger}_{k,s} \, \Theta^*_{k,s} e^{-\frac{i}{\hbar} \underleftrightarrow{S} \overleftrightarrow{\Phi}}\right],
\end{eqnarray}
where $\hat{S}^{\alpha}=s \,\hat{\gamma}^{\alpha} $ is the spin operator, and
\begin{equation}
\frac{\delta }{\hat{\delta\Phi}^{\alpha}}\left(\underleftrightarrow{S} \overleftrightarrow{\Phi}\right) =  \left(2 g_{\alpha\beta}  \mathbb{I}_{4\times 4} - \hat{\gamma}_{\alpha} \hat{\gamma}_{\beta} \right) \hat{S}^{\beta} = 2 \hat{S}_{\alpha} - \hat{\gamma}_{\alpha} \,s,
\end{equation}
such that $s\,\mathbb{I}_{4\times 4}= \frac{1}{4} \hat{S}_{\beta} \hat{\gamma}^{\beta}$.

\section{Quantum dynamics of $\hat{\Psi}_{\alpha}$ on the extended manifold}\label{qd}

The boundary terms in (\ref{delta}), on the extended manifold described by the connections components (\ref{conn}), are
given by (\ref{bor}), where $g^{\alpha\beta} \delta R_{\alpha\beta} = \Theta(x^{\alpha}) = {\Lambda}\, g^{\alpha\beta} \delta
g_{\alpha\beta}$ is the flux of the 4-vector $\hat{\delta W}^{\alpha}$ that cross any $3D$ closed manifold defined on an arbitrary region of the background manifold, which is considered as Riemannian and is characterized by the Levi-Civita connections. Here, we are considering  the covariant derivative of the metric tensor and some vector on the extended manifold, with self-interactions included, which are respectively given by
\begin{eqnarray}
\hat{g}_{\beta\alpha\|\gamma}& =&\nabla_{\gamma} g_{\beta\alpha} -\epsilon\, \left(g_{\beta\gamma} \hat{\Psi}_{\alpha} + \hat{\Psi}_{\beta} g_{\alpha\gamma} \right) + 2\left(1-\xi^2\right) \hat{\Psi}_{\gamma} \, g_{\alpha\beta}, \label{gg} \\
\left[\Upsilon^{\alpha}\right]_{||\beta}&=& \nabla_{\beta}\Upsilon^{\alpha} + \delta\Gamma^{\alpha}_{\epsilon\beta}\Upsilon^{\epsilon} - (1-\xi^2) \Upsilon^{\alpha}\hat\Psi_{\beta}
= \nabla_{\beta}\Upsilon^{\alpha} + \left(\epsilon \,\hat{\Psi}^{\alpha}\Upsilon_{\beta}-\Upsilon^{\alpha}\hat{\Psi}_{\beta}\right) + \xi^2 \Upsilon^{\alpha}\hat\Psi_{\beta},
\end{eqnarray}
where $\xi$ is the self-interaction constant, $\nabla_{\gamma} g_{\beta\alpha}=0$ is the covariant derivative on the Riemann manifold, and $\|$ denotes the covariant derivative on the extended manifold. If we deal with fermions, creation and destruction operators must comply\cite{ab}
\begin{eqnarray}
\frac{4 s^2\,   L^2_p}{\hbar^2} \left(|A_{k,s}|^2 + |B_{k,s}|^2\right) &=&  \, \left(\frac{c^3 M^3_p}{\hbar}\right)^2. \label{q2}
\end{eqnarray}
The expectation value for the local particle-number operator for bosons with wave-number norm $k$ and spin $s$, $\hat{N}_{k,s}$, is given by
\begin{equation}\label{np}
\left<B\left|\hat{N}_{k,s}\right|B\right> =n_{k,s}\,\left(\frac{\hbar}{c^3 M^3_p}\right)^2\,\int d^4x \sqrt{-g} \int d^4\phi \, \left< B\left| \left\{\hat{\slashed{\Psi}}({\bf x}, {\bf \phi}), \hat{\slashed{\Psi}}^{\dagger}({\bf x}, {\bf \phi}) \right\}\right| B \right> =n_{k,s} \,\mathbb{I}_{4\times4},
\end{equation}
where the slashed spinor fields are: $\hat{\slashed{\Psi}}=\bar{\gamma}^{\mu} {\hat\Psi}_{\mu}$, $\hat{\slashed{\Psi}}^{\dagger} = \left(\bar{\gamma}^{\mu} {\hat\Psi}_{\mu}\right)^{\dagger}$. Furthermore, these fields comply with the algebra
\begin{equation}
\left< B\left| \left\{\hat{\slashed{\Psi}}({\bf x}, {\bf \phi}), \hat{\slashed{\Psi}}^{\dagger}({\bf x}', {\bf \phi}') \right\}\right|B \right>
 = \frac{4 s^2 \,L^2_p}{\hbar^2} \left(|A_{k,s}|^2 + |B_{k,s}|^2\right) \,\,  \sqrt{\frac{\eta}{g}} \,\,\delta^{(4)} \left({\bf x} - {\bf x}'\right) \,\delta^{(4)} \left({\bf \phi} - {\bf \phi}'\right).
\end{equation}
This result is valid in any relativistic scenario. To connect the Fock-space theory and the ordinary quantum mechanics one can introduce the wave function in position space by using the definition of a kind of $n_{k,s}$-particle state vector that describes a system of $n_{k,s}$ particles that are localized in coordinate space at the points ${\bf x}_1; {\bf \phi}_1...{\bf x}_n; {\bf \phi}_n$:
\begin{displaymath}
\left|{\bf x}_1,{\bf x}_2,...,{\bf x}_n;{\bf \phi}_1,{\bf \phi}_2,...,{\bf \phi}_n \right> =\prod^{ord}_{k,s} \,\frac{1}{\sqrt{n_{k,s}!}} \hat{\slashed{\Psi}}^{\dagger}({\bf x}_1; {\bf \phi}_1)...\hat{\slashed{\Psi}}^{\dagger}({\bf x}_n; {\bf \phi}_n)\left|B\right>,
\end{displaymath}
where here $\left|B\right>$ is our reference state. The reference state $\left|B\right>$, is not a vacuum state, but describes the Riemannian (classical) reference with respect to which we describe the quantum system on a curved background state. Notice the order prescription in the product: $\prod^{ord}_{k,s}$. This fact is because we are dealing with fermions and the Pauli's exclusion principle exclude the possibility that two particles can stay in the same state.

In order to obtain the dynamic equations of quantum spinor fields on the extended manifold, we must variate the Ricci tensor using the Palatini identity\cite{pal}, $\hat{\delta{R}}^{\alpha}_{\beta\gamma\alpha}=\hat{\delta{R}}_{\beta\gamma}= \left(\hat{\delta\Gamma}^{\alpha}_{\beta\alpha} \right)_{\| \gamma} - \left(\hat{\delta\Gamma}^{\alpha}_{\beta\gamma} \right)_{\| \alpha}$\footnote{To calculate the Ricci tensor we must develop the following covariant derivatives:\\
\begin{eqnarray*}
\left(\hat{\delta\Gamma}^{\alpha}_{\beta\alpha}\right)_{\| \gamma} & = & \epsilon \,\left. \left( \hat{\Psi}^{\alpha}\, g_{\beta\alpha} \right)\right|_{\|\gamma}=
\epsilon \,\hat{\Psi}_{\beta\|\gamma},  \\
\left(\hat{\delta\Gamma}^{\alpha}_{\beta\gamma} \right)_{\| \alpha} & = & \epsilon \left. \left( \hat{\Psi}^{\alpha}\,g_{\beta\gamma}\right)\right|_{\|\alpha}= \epsilon\,\left[\hat{\Psi}^{\alpha}_{\,\,\,\,\|\alpha}\,g_{\beta\gamma} + \hat{\Psi}^{\alpha} g_{\beta\gamma\|\alpha}\right],
\end{eqnarray*}
where
\begin{eqnarray*}
\hat{\Psi}_{\beta\|\gamma} & = & \nabla_{\gamma} \hat{\Psi}_{\beta} - \epsilon \,g_{\beta\gamma}\,\hat{\Psi}^{\mu}\hat{\Psi}_{\mu} +(1-\xi^2)\,\hat{\Psi}_{\beta}\hat{\Psi}_{\gamma}, \\
\hat{\Psi}^{\alpha}_{\,\,\,\,\|\alpha} & = & \nabla_{\alpha} \hat{\Psi}^{\alpha} + \xi^2\, \hat{\Psi}^{\alpha} \hat{\Psi}_{\alpha} , \\
g_{\beta\gamma\|\alpha} &=& -\epsilon \left(g_{\alpha\gamma} \hat{\Psi}_{\beta} + g_{\gamma\beta} \,\hat{\Psi}^{\alpha} \right) + 2(1-\xi^2)\,
\hat{\Psi}_{\alpha} \,g_{\beta\gamma}.
\end{eqnarray*}
}, is
\begin{equation}
\hat{\delta{R}}_{\beta\gamma} = \epsilon\left[\nabla_{\gamma} \hat{\Psi}_{\beta} - \left(\epsilon+1-\xi^2\right) g_{\beta\gamma} \left(\hat{\Psi}^{\nu} \hat{\Psi}_{\nu}\right)
-g_{\beta\gamma} \left(\nabla_{\nu} \hat{\Psi}^{\nu}\right) +\left(\epsilon+1-\xi^2\right)\hat{\Psi}_{\beta}\hat{\Psi}_{\gamma}\right]. \label{ricci}
\end{equation}
This tensor has both, symmetric $\hat{U}_{\beta\gamma}$ and antisymmetric  $\hat{V}_{\beta\gamma}$ contributions:
\begin{eqnarray}
\hat{U}_{\beta\gamma}&=&\frac{\epsilon}{2} \left( \nabla_{\beta} \hat{\Psi}_{\gamma}+\nabla_{\gamma} \hat{\Psi}_{\beta}\right) \nonumber \\ &-&\epsilon\left[\left(\epsilon+1-\xi^2\right) g_{\beta\gamma} \left(\hat{\Psi}^{\nu} \hat{\Psi}_{\nu}\right)
+g_{\beta\gamma} \left(\nabla_{\nu} \hat{\Psi}^{\nu}\right)-\frac{1}{2}\left(\epsilon+1-\xi^2\right)\left\{\hat{\Psi}_{\beta},\hat{\Psi}_{\gamma}\right\}\right], \label{u} \\
\hat{V}_{\beta\gamma} &=& \frac{\epsilon}{2} \left( \nabla_{\gamma} \hat{\Psi}_{\beta} -\nabla_{\beta} \hat{\Psi}_{\gamma} \right)+\frac{\epsilon}{2}\left(\epsilon+1-\xi^2\right)\left[\hat{\Psi}_{\beta}, \hat{\Psi}_{\gamma}\right] . \label{v}
\end{eqnarray}
Furthermore, the purely antisymmetric tensor $\hat{\delta{R}}^{\alpha}_{\alpha\beta\gamma}\equiv \hat{\Sigma}_{\beta\gamma}$, is
\begin{equation}\label{sigma}
\hat{\Sigma}_{\beta\gamma} = \epsilon\left( \nabla_{\gamma} \hat{\Psi}_{\beta}-\nabla_{\beta} \hat{\Psi}_{\gamma}\right) +\epsilon\left(1-\xi^2\right) \left[ \hat{\Psi}_{\gamma}, \hat{\Psi}_{\beta} \right],
\end{equation}
which, once we set $\epsilon\left(1-\xi^2\right)=i\,{\bf g}$, {\it is the Yang-Mills tensor} on the curved background manifold and describes for example the gluonic (strong) interactions of quarks. Notice that (\ref{sigma}) can describe the gluon field strength tensor related to the gluon field components: $\hat{\Psi}^{\mu} = \frac{\lambda^n}{2}\, \hat{A}^{\mu}_n $, such that  $\lambda^n $ are the eight ($3\times 3$) Gell-Mann matrices in the $SU(3)$ group representation and ${\bf g}=\sqrt{4\pi \alpha_s}$, $\alpha_s$ being the coupling constant of the strong force. The experimental value is $\alpha_s\simeq 0.1182$\cite{pdb}, that corresponds to $\epsilon(1-\xi^2)=1.218746\,i)$.

The Einstein tensor on the extended manifold can be defined using the symmetric contribution of the Ricci tensor (\ref{ricci}): $\hat{\delta{G}}_{\beta\gamma}= \hat{U}_{\beta\gamma} - \frac{1}{2} g_{\beta\gamma} \hat{U}$, with $\hat{U}=g^{\alpha\beta} \hat{U}_{\alpha\beta}$:
\begin{equation}
\hat{\delta{G}}_{\beta\gamma}=\frac{\epsilon}{2} \left[\left( \nabla_{\beta} \hat{\Psi}_{\gamma}+\nabla_{\gamma} \hat{\Psi}_{\beta}\right)+ g_{\beta\gamma}
\nabla_{\nu} \hat{\Psi}^{\nu}+\left(\epsilon+1-\xi^2\right)\left\{\hat{\Psi}_{\beta},\hat{\Psi}_{\gamma}\right\}-2\left(\epsilon+1-\xi^2\right) g_{\beta\gamma} \left(\hat{\Psi}^{\nu} \hat{\Psi}_{\nu}\right)\right]. \label{Gab}
\end{equation}
Therefore, the operatorial Einstein equation (\ref{tran2}), with (\ref{Gab}) and (\ref{ff}) , results to be
\begin{eqnarray}
&& \frac{\epsilon}{2} \left[ \left( \nabla_{\beta} \hat{\Psi}_{\gamma}+\nabla_{\gamma} \hat{\Psi}_{\beta}\right)+ g_{\beta\gamma}
\nabla_{\nu} \hat{\Psi}^{\nu}+\left(\epsilon+1-\xi^2\right)\left\{\hat{\Psi}_{\beta},\hat{\Psi}_{\gamma}\right\}-2\left(\epsilon+1-\xi^2\right) g_{\beta\gamma} \left(\hat{\Psi}^{\nu} \hat{\Psi}_{\nu}\right)\right]\nonumber \\
&=&-\frac{1}{2} \left[ \left(\nabla_{\beta}\hat{\Psi}_{\gamma}+\nabla_{\gamma} \hat{\Psi}_{\beta}\right) + m\, \left\{\hat{\Psi}_{\beta}, \hat{\Psi}_{\gamma} \right\}\right]. \label{eii}
\end{eqnarray}

In order to describe massless bosons and charged bosons, we define
$\hat{\cal{N}}_{\beta\gamma} = {1\over 2} \hat{V}_{\beta\gamma} + {1\over 4}\hat{{\Sigma}}_{\beta\gamma} $ and $\hat{\cal{M}}_{\beta\gamma} = {1\over 2} \hat{V}_{\beta\gamma} - {1\over 4} \hat{{\Sigma}}_{\beta\gamma} $
\begin{eqnarray}
\hat{\cal{N}}_{\beta\gamma} & = & \frac{\epsilon}{2} \left(\nabla_{\gamma} \hat{\Psi}_{\beta} -\nabla_{\beta} \hat{\Psi}_{\gamma}\right) + \frac{\epsilon^{2}}{4}\left[ \hat{\Psi}_{\beta}, \hat{\Psi}_{\gamma}\right], \label{n1}\\
\hat{\cal{M}}_{\beta\gamma} & = & \left(\frac{\epsilon}{2}(1-\xi^2)-\frac{\epsilon^2}{4}\right)  \left[ \hat{\Psi}_{\beta}, \hat{\Psi}_{\gamma}\right]. \label{n2}
\end{eqnarray}
The symmetric tensor $\hat{\delta{G}}_{\beta\gamma}$, with the antisymmetric ones $\hat{{\Sigma}}_{\beta\gamma}$, $\hat{\cal{N}}_{\beta\gamma}$ and $\hat{\cal{M}}_{\beta\gamma}$, describe all the possible interactions of fermions and bosons on the extended manifold. On one hand, we see that $\hat{\cal{N}}_{\beta\gamma}$ corresponds to a Yang-Mills strength tensor, and describes massless bosons like photons or gravitons with self-interactions included. On the other hand $\hat{{\Sigma}}^{\beta\gamma}$ describes bosons which interact with colour charge (like gluons). We can see that $\hat{\cal{N}}_{\beta\gamma}$ is independent of the coupling and all the coupling-related information is on the tensor $\hat{\cal{M}}_{\beta\gamma}$. Also, $\hat{\cal{M}}_{\beta\gamma}$ does not contain a free-like term, so that can describe massive and charged vector bosons (may be $X^{\pm}$ and $Y$ mediators predicted by GUT theories) which will result in a non-wave-like propagating field dynamics. If the four fundamental tensors are conserved on the extended manifold, then they must comply
\begin{equation}
\left(\hat{\delta{G}}^{\beta\gamma}\right)_{\|\gamma} =0, \quad \left(\hat{\cal{N}}^{\beta\gamma}\right)_{\|\gamma}=0, \quad \left(\hat{{\Sigma}}^{\beta\gamma}\right)_{\|\gamma}=0 , \quad     \left(\hat{\cal{M}}^{\beta\gamma}\right)_{\|\gamma}=0.
\end{equation}

\section{de Sitter expansion: fermionic origin of cosmological constant}\label{ex}

To explore the model we shall consider a de Sitter (inflationary) expansion, where the background spacetime is described by the line element
\begin{equation}\label{m}
dx^2 = a^2(\eta) \left[ d\eta^2 - \delta_{ij} \,dx^i dx^j\right],
\end{equation}
where $\eta$, that runs from $-\infty$ to zero, is the conformal time of the universe, which is considered as spatially flat, isotropic and
homogeneous.

\subsection{Background dynamics}

In a de Sitter expansion the scale factor of the universe is $a(\eta) =-{1\over H\,\eta}$. If the expansion is governed by the inflaton field $\varphi$, and it is non-minimally coupled to gravity, the universe can be described by the action
\begin{equation}
{\cal I} = \int \,d^4x\,\sqrt{-g}\,\left[\frac{{\cal R}}{2\kappa} + {\cal L}_{\varphi}\right],
\end{equation}
where ${\cal L}_{\varphi}=-\left[{1\over 2} (\varphi')^2 -V(\varphi)\right]$, where the scalar potential is a constant $V(\varphi)= {3 H^2\over \kappa}$ and the {\it prime} denotes the derivative with respect to $\eta$. Furthermore, $H=(a'/a)$ is the Hubble parameter which remains constant during the de Sitter expansion\cite{bcms}. The background inflaton field has the equation of motion
\begin{equation}
\varphi''+ 2 H \varphi' + \frac{\delta V}{\delta \varphi} =0,
\end{equation}
with solution $\varphi(\eta)=\varphi_0$. Therefore, the kinetic component of ${\cal L}_{\varphi}$ is zero.

\subsection{Calculation of $\Lambda$ from fermion fields}

We are aimed to calculate $\Lambda$ in a de Sitter expansion originated by the flux through the 3D-Gaussian hypersurface, described by (\ref{bor}), that becomes from $ \left\{\hat{\Psi}_{\alpha}, \hat{\Psi}_{\beta} \right\}$, which has an exclusively fermionic contribution. To describe the system we shall consider a Lagrangian density for fermion fields with mass $m$, described by
\begin{equation}\label{fl}
\hat{\cal L} =\frac{g^{\beta\gamma}}{4 \kappa}\,\left[\left(\nabla_{\beta}\hat{\Psi}_{\gamma}+\nabla_{\gamma} \hat{\Psi}_{\beta}\right)+  \frac{m}{2} \left\{\hat{\Psi}_{\beta}, \hat{\Psi}_{\gamma} \right\}\right],
\end{equation}
such that, using the expression (\ref{t2}), we obtain the stress tensor related to this Lagrangian
\begin{equation}\label{ff}
\kappa\,\hat{\delta T}_{\mu\nu} =\frac{1}{2} \left[ \left(\nabla_{\mu}\hat{\Psi}_{\nu}+\nabla_{\nu} \hat{\Psi}_{\mu}\right) + m\, \left\{\hat{\Psi}_{\mu}, \hat{\Psi}_{\nu} \right\}\right].
\end{equation}

The matrices that generates the spacetime $g_{\mu\nu} \,\mathbb{I}_{4\times 4} = (1/2) \left\{\hat{\gamma}_{\mu}, \hat{\gamma}_{\nu}\right\}$, are:
\begin{equation}
\hat{\gamma}^{\mu} = a^2(\eta) \,\gamma^{\mu},
\end{equation}
with the $\gamma^{\mu}$ given by (\ref{gamm}).

Using the fact that $\hat{\delta G}=-\hat{\delta R}\equiv -\hat{U} = -g^{\alpha\beta} \,\hat{U}_{\beta\gamma} = -4 \hat{\Lambda}=-\kappa\, \hat{\delta T}$, and by contracting the quantum Einstein equations (\ref{eii})
\begin{equation}
\nabla_{\nu} \, \hat{\Psi}^{\nu} = \frac{m}{2(3\epsilon -1)} g^{\mu\nu}\,\left\{\hat{\Psi}_{\mu}, \hat{\Psi}_{\nu}\right\},
\end{equation}
we obtain that the cosmological parameter $\Lambda $, is
(\ref{lam}), is
\begin{equation}
\Lambda = \frac{3m}{8} \left(\frac{\epsilon}{3\epsilon-1}\right) g^{\mu\nu} \int\, dv'\,  \left<B\left|\left\{\hat{\Psi}_{\mu}, \hat{\Psi}_{\nu}\right\}\right|B\right>,
\end{equation}
where it is easy to see that $\Lambda$ is originated by quantum self-interaction contributions of primordial coherent fermionic fields, that comply with the algebra (\ref{f}), and $dv'=\sqrt{-g}\, d^4\phi'\,d^4x'=a^3(\tau)\, d^4\phi'\,d^4x'$. The fermion fields obey the quantization rules on a de Sitter space-time
\begin{eqnarray}
&& \left< B\left| \left\{\hat{\Psi}_{\mu}({\bf x}, {\bf \phi}), \hat{\Psi}_{\nu}({\bf x}', {\bf \phi}') \right\}\right|B \right> = \frac{1}{2(c\tau)^2}\,\,\frac{s^2}{\hbar^2}\, \left\{\hat{\gamma}_{\mu}, \hat{\gamma}_{\nu} \right\}\,\mathbb{I}_{4\times 4} \,  \sqrt{\frac{\eta}{g}}  \,\,\delta^{(4)} \left({\bf x} - {\bf x}'\right) \,\delta^{(4)} \left({\bf \phi} - {\bf \phi}'\right),\label{f}
\end{eqnarray}
where $c\tau$ is the size of the causal radius, $c$ is the light velocity in the vacuum, $\tau$ is same characteristic time-scale of the self-interaction, such that in a cosmological framework we could take $\tau=1/H$, where $H$ is the Hubble parameter. In this fashion the theory includes the effects of a quantum field in the geometry of spacetime. Therefore, we obtain
\begin{equation}
 g^{\mu\nu} \int\, dv'\,  \left<B\left|\left\{\hat{\Psi}_{\mu},\hat{\Psi}_{\nu}\right\}\right|B\right> = \frac{4}{(c\tau)^2} \frac{s^2}{\hbar^2} \,\mathbb{I}_{4\times 4},
\end{equation}
and therefore, for fermion fields with $s=\hbar/2$, we obtain
\begin{equation}
\Lambda = \frac{m}{8} \left(\frac{3\epsilon}{3\epsilon-1}\right) \frac{H^2}{c^2} \,\mathbb{I}_{4\times 4},
\end{equation}
which is
\begin{equation}
\Lambda =  \frac{3 H^2}{c^2} \,\mathbb{I}_{4\times 4},
\end{equation}
for
\begin{equation}\label{mas}
m= 8\,\left(\frac{3\epsilon-1}{\epsilon}\right) \geq 0.
\end{equation}
This is an important result that say us that, when the geometrical field $\hat{\delta W}^{\alpha}$ is equal than the fermion field $\hat{\Psi}^{\alpha}$, the flux $\hat{\Theta}$ is integrable in terms of both: $\hat{\Psi}^{\alpha}$ and $\hat{\delta W}^{\alpha}$. This fact holds when $\epsilon=1/3$ and hence $m=0$. However, when $\epsilon > 1/3$, the flux $\hat{\Theta}$ is integrable in terms of $\hat{\delta W}^{\alpha}$, but not of $\hat{\Psi}^{\alpha}$. In this case fermion fields $\hat{\Psi}^{\alpha}$ with $s=\hbar/2$, acquire a mass given by (\ref{mas}), which is the responsible for the existence of $\hat{\Lambda}$. Notice that for the result (\ref{mas}), one obtains the asymptotic value for $m$:
\begin{equation}
\left. m \right|_{\epsilon \rightarrow \infty}\rightarrow 24.
\end{equation}
The observational present day value for $\Lambda$ is\cite{pdb}:
\begin{equation}
\Lambda = 1.58908742 \times 10^{-52} \quad {\rm meters}^{-2}.
\end{equation}
which is obtained for an Hubble constant $h=0.678$ and a Hubble radius $(c/H)=1.374 \times\,10^{26}$ meters.

\section{Final comments}\label{conc}

We have studied an USF theory that includes both, a quantum description of spacetime and a relativistic description of spinor fields which describes bosons and fermions on curved background spacetimes described by a Riemann manifold. To make it, we have used as generators of the $4$-quantum-operators the $4\times4$-$\gamma^{\mu}$ matrices, which comply with the Clifford algebra, and assures that all operators do that. In this framework, the extended Ricci tensor has symmetric and antisymmetric contributions, but not as a consequence of the torsion, rather originated by the structure coefficients induced by the Clifford algebra of the generators $\gamma^{\mu}$, with we construct the operators and the spacetime. The spacetime has $8$ dimensions, $4$ dimensions coordinates and $4$ that describes the inner space $\phi^{\mu}$ with canonical moments $s^{\mu}$ that describes the components of spin.

All the quantum spinor fields become from the boundary terms in (\ref{delta}), on the extended manifold described by the connections components (\ref{conn}). The flux through the $3D$-gaussian hypersurface is
given by (\ref{bor}), where $g^{\alpha\beta} \hat{\delta R}_{\alpha\beta} = \hat{\Theta}(x^{\alpha}|\phi^{\alpha}) = \hat{\Lambda}\, g^{\alpha\beta} {\delta
g}_{\alpha\beta}$ is the flux of the 4-vector $\hat{\delta W}^{\alpha}= \hat{\delta\Gamma}^{\epsilon}_{\beta\epsilon} g^{\beta\alpha}- \hat{\delta \Gamma}^{\alpha}_{\beta\gamma} g^{\beta\gamma}$ that cross any $3D$ closed manifold defined on an arbitrary region of the background Riemann manifold. In order to $\delta{\cal I}=0$ in (\ref{delta}), we must require that the extended manifold be an Einstein's one:
\begin{displaymath}
\delta R_{\alpha\beta} = \Lambda\,\delta g_{\alpha\beta},
\end{displaymath}
so that, because $\delta g^{\alpha\beta}\,g_{\alpha\beta}=-\delta g_{\alpha\beta}\,g^{\alpha\beta}$, we obtain the extended Einstein's equations (\ref{e1}). As we have shown in Sect. (\ref{cos}), this implies that the "cosmological constant" comes from the extended Einstein tensor: $\Lambda = -{1\over 4} \delta G={1\over 4} \delta R={\kappa\over 4} \delta T$. Of course, in general $\Lambda$ is not a constant on the extended manifold, but in the example of a de Sitter expansion, which was studied in this paper, it is so. We have calculated its value, which is determined exclusively by massive fermion fields. Of course this is explained only by an entanglement of quantum fermions (with spin $1/2$), which are coherent at the beginning of the expansion of the universe. They should constitute the large-scale dark energy in the universe.

\section*{Acknowledgements}

\noindent The authors acknowledge CONICET, Argentina (PIP 11220150100072CO) and UNMdP (EXA852/18) for financial support.

\end{document}